\documentclass[aps,prl,twocolumn,superscriptaddress,10pt]{revtex4-1}

\usepackage{overpic}
\usepackage{physics}
\usepackage{graphicx}
\usepackage[caption=false]{subfig}

\newcommand{\nonmark}{\mathcal{N}}
\newcommand{\liou}{\mathcal{L}}

\begin{document}

\title{Phenomenological measure of quantum non-Markovianity}
\author{Adam Winick}
\affiliation{Institute for Quantum Computing and Department of Applied Mathematics, University of Waterloo, Waterloo, Canada}

\author{Joel J. Wallman}
\affiliation{Institute for Quantum Computing and Department of Applied Mathematics, University of Waterloo, Waterloo, Canada}

\author{Joseph Emerson}
\affiliation{Institute for Quantum Computing and Department of Applied Mathematics, University of Waterloo, Waterloo, Canada}
\affiliation{Canadian Institute for Advanced Research, Toronto, Ontario M5G 1Z8, Canada}

\begin{abstract}
Non-Markovian effects are ubiquitous in physical quantum systems and remain a significant challenge to achieving high-quality control and reliable quantum computation, but due to their inherent complexity, are rarely characterized. Past approaches to quantifying non-Markovianity have concentrated on small, simple systems, and we show that such measures can capture properties that are irrelevant to applications. With this insight, we propose a method for constructing relevant quantifiers of non-Markovian dynamics and illustrate the scheme's computability by characterizing a model quantum dot array.
\end{abstract}

\maketitle

Dynamical semigroups are often applied to describe the interaction of a quantum system with its environment. The corresponding Lindblad equation \cite{Lindblad1976,Gorini1976} depicts memoryless dynamics and an irreversible loss of information through decoherence mechanisms. However, in many settings, the assumptions that justify a semigroup fail, and systems exhibit memory effects that give rise to a resurgence of coherence and information flow back into the system. These memory effects are the principal characteristics of non-Markovian dynamics in quantum systems.

Identifying non-Markovian effects and their respective magnitudes and timescales is critical to the success of quantum computation \cite{Bylicka2014}, where such effects can have a significant impact on the performance of fault tolerant quantum error correction \cite{Terhal2005,Aharonov2006} and the robustness of error characterization methods such as randomized benchmarking \cite{Emerson2005,Epstein2014,Wallman2015c,Fogarty2015,Yang2018}. Characterizing these memory effects is an essential first step towards their suppression. Non-Markovian effects also play a role in, for example, quantum biology \cite{Rebentrost2011}, quantum key distribution \cite{Vasile2011},  and quantum metrology \cite{Chin2012}. Moreover, recent results show that the manipulation of reservoir spectral properties can improve quantum control \cite{Verstraete2009,Wineland2000}, and hence characterizing systems beyond Lindblad master equations is the pathway to unlock these capabilities.

Quantifying non-Markovian processes in quantum systems has been a standing challenge for several decades. Recent work has resolved the theoretical problem of describing non-Markovian processes that reduces to its classical counterpart in an appropriate limit \cite{Pollock2018a,Pollock2018b}. The required experimental resources become practically insurmountable for complex multipartite systems, and there is a need for so-called measures of non-Markovianity. Several approaches have been proposed to construct measures of non-Markovianity based on the geometry of states \cite{Wolf2008a,Wolf2008b}, the violation of CP-divisibility \cite{Rivas2010,Chruscinski2011}, monotonicity under CP maps \cite{Breuer2009,Laine2010,Vasile2011,Luo2012,Bylicka2014,Lu2011,Lorenzo2013}, and other principles \cite{Chruscinski2014,Hall2014}. To date, all proposed measures require full process tomography which is unrealistic for even moderately sized systems of a few qubits. All but one of the measures above further require an optimization that becomes intractable for systems beyond one or two qubits. The remaining measure \cite{Rivas2010} assumes knowledge of the dynamical map or an explicit form of the master equation and is therefore not measurable for complex system-environment couplings and internal environment dynamics.

In this Letter, we introduce a new approach to quantify non-Markovianity that is directly related to how it manifests under common metrics of interest. We begin with a review of the interpretation and measure of the degree of non-Markovianity proposed by \citet{Breuer2009}, which we denote BLP below. By considering the behavior of the measure for a toy model, we highlight some of its limitations. We then use this insight to define our method. A numerical experiment demonstrates that our technique can be used to gauge the effect of non-Markovian noise on systems that are orders of magnitude larger than those analyzed in previous studies.


A key concept for our approach is that memory effects that appear in non-Markovian dynamics arise from an exchange of information between a system and its environment. The exchange is recognizable in several suitable quantities, e.g., quantum relative entropy, but for concreteness, we consider the trace distance $D(\rho_1, \rho_2)$  between quantum states $\rho_1$ and $\rho_2$. Among the many properties of the trace distance, it is an operationally meaningful quantifier of the distinguishability between two states. Suppose Alice prepares a system in either the state $\rho_1$ or $\rho_2$, each with probability $\frac{1}{2}$ and sends it to Bob who performs a measurement to identify if the transmitted state was $\rho_1$ or $\rho_2$. It can be shown that with an optimal measurement, Bob can successfully identify the state with probability $\frac{1}{2}\left[ 1 + D(\rho_1, \rho_2) \right]$.

It is well-known that there is no quantum operation, defined on the system Hilbert space, that can increase the trace distance between states. More precisely, a completely positive non-trace-increasing map $\Phi$ is a contraction on the trace distance metric,
\begin{equation} \label{eqn:CPmapineq}
    D(\Phi \rho_1, \Phi \rho_2) \leq D(\rho_1, \rho_2) \,.
\end{equation}
Consider a Markovian master equation for the reduced state $\rho$ of an open system characterized by the Lindblad generator $\liou$. The solution is a dynamical semigroup $\Phi(t)=\exp(\liou t)$. As a consequence of the semigroup property $\Phi(\tau + t) = \Phi(\tau)\Phi(t)$ for all $\tau, t \geq 0$,
\begin{equation} \label{eqn:Dynamicalineq}
    D(\Phi(t + \tau)\rho_1, \Phi(t + \tau)\rho_2) \leq D(\Phi(t)\rho_1, \Phi(t)\rho_2) \,.
\end{equation}
The inequality holds for the larger class of positivity preserving maps. For example, a time-inhomogeneous Lindblad generator $\liou(t)$ describes time-dependent Markovian processes where the map $\Phi(t)$ need not be a dynamical semigroup. If instead we define a two-parameter family \cite{Breuer2009} of dynamical maps $\Phi(t_2, t_1)$ where $\Phi(t, 0) = \Phi(t)$, we obtain the comparable semigroup property $\Phi(t + \tau, 0) = \Phi(t + \tau, \tau)\Phi(\tau, 0)$.

From these properties of the trace distance, \eqref{eqn:Dynamicalineq} implies that Markovian processes lead to the unidirectional flow of information from a system to an environment. In contrast, a non-Markovian process might lead to an intermittent reversal in the direction of information flow and an improved distinguishability of states. The information flux between two states $\rho_{1,2}$ is
\begin{equation}
    \sigma(t, \rho_{1,2}) = \frac{d}{dt}D(\Phi(t)\rho_1,\Phi(t)\rho_2) \,.
\end{equation}
At a time $t > 0$, if $\sigma(t, \rho_{1,2}) < 0$, the distinguishability between the pair of states is decreasing and information flows from the system. The interpretation motivates the BLP measure of non-Markovianity,
\begin{equation} \label{eqn:BLP}
\nonmark_\text{BLP}(\Phi) = \sup_{\rho_{1,2}} \int_{\sigma > 0} dt \sigma(t, \rho_{1,2}) \,,
\end{equation}
where the supremum is taken over all mixed states.

Because of the maximization inherent to the measure, for this model it is independent of $N$! Next, suppose that we then apply a strongly decohering Markovian process to the $N$ uncoupled qubits over some time. Again, the BLP measure remains constant. An immediate consequence is that if one only looks at the BLP measure in the limit $N \rightarrow \infty$, then one concludes that a process is highly non-Markovian when actually, except with vanishing probability, every state evolves under an effective Markovian channel!

In a generic quantum computing problem, we select some initial state $\rho$ and apply a noisy operation $\Phi$. If the process has `concentrated' non-Markovianity, like in the preceding example, the BLP measure can be inaccurate when trying to understand the degree to which non-Markovianity discernably affects that overall system.  The imprecision is unsurprising for an optimal state pair $\rho_{1,2}$ need not relate to the set $\mathcal{S}$ of valid initial states for the computing problem.

Building upon the information flow description of non-Markovianity, what aspect of that flow is discernable in practice? Rather than probing the distinguishability between an optimal state pair, we examine the expected distinguishability between states in $\mathcal{S}$. Without an a priori set, we focus on the expected distinguishability of all pure states and introduce the average pure state distinguishability,
\begin{equation} \label{eqn:stateDist}
    D_\text{avg}(\Phi) = \iint d\psi_{1,2} D(\Phi\psi_1, \Phi\psi_2) \,,
\end{equation}
where $d\psi_{1,2}$ denotes the natural invariant Haar measure.

At a time $t$, if $\partial_t D_\text{avg} < 0$, then the expected distinguishability between any pair of states is decreasing and information flows only out from the system. There is an opposite interpretation of $\partial_t D_\text{avg} > 0$.

We define the \emph{net} flux,
\begin{equation} \label{eqn:NetFlow}
    \sigma_\text{avg}(t) = \frac{\partial D_\text{avg}}{\partial t} = \iint d\psi_{1,2} \sigma(t, \psi_{1,2}) \,,
\end{equation}
which we can decompose into its positive and negative contributions,
\begin{align} \label{eqn:DecomposeFlow}
    \sigma_\text{avg}(t) &= \iint_{\sigma > 0} d\psi_{1,2} \sigma(t, \psi_{1,2}) +  \iint_{\sigma < 0} d\psi_{1,2} \sigma(t, \psi_{1,2}) \nonumber \\
    &= \sigma_+(t) + \sigma_-(t) \,.
\end{align}
In the above, we implicitly defined $\sigma_+$ and $\sigma_-$ to denote the overall strength of  non-Markovian and Markovian like processes respectively. According to these measures, a process is non-Markovian at a time $t$ if $\sigma_+ > 0$  and a process is \emph{purely} non-Markovian at a time $t$ if $\sigma_+ > 0$ and $\sigma_- = 0$. We call a process \emph{strongly} non-Markovian at a time $t$ if $\sigma_+ > \sigma_-$. 

By integrating the information flux, we arrive at two possible measures of the strength of non-Markovianity: a measure of average non-Markovianity,
\begin{equation} \label{eqn:avgNM}
    \nonmark_\text{avg}(\Phi) = \int_{\sigma_\text{avg} > 0}dt \sigma_\text{avg}(t) \,,
\end{equation}
and of pure non-Markovianity,
\begin{equation} \label{eqn:pureNM}
    \nonmark_p(\Phi) = \int dt \sigma_+(t) \,.
\end{equation}

To illustrate the significance of the new measures, we revisit the $N+1$ qubit toy model. While the BLP measure was constant, with only the $ZZ$ coupling, both measures tend to zero like $1/N$ as $N\to\infty$. Instead of describing extremal behavior these measures relate to typical characteristics. Now suppose we gradually turn on a strongly decohering Markovian channel that acts on the $N$ uncoupled qubits. Both the BLP and pure measure remain constant, and the average measure decreases. Thus by examining $\nonmark_{\text{avg}, p}$, we can identify the common non-Markovianity qualities of a process. In general, the BLP measure is a lower bound on our measures with
\begin{equation}
    \nonmark_\text{avg}(\Phi) \leq \nonmark_p(\Phi) \leq  \nonmark_\text{BLP}(\Phi) \,,
\end{equation}
where the leftmost inequality follows from Jensen's inequality.



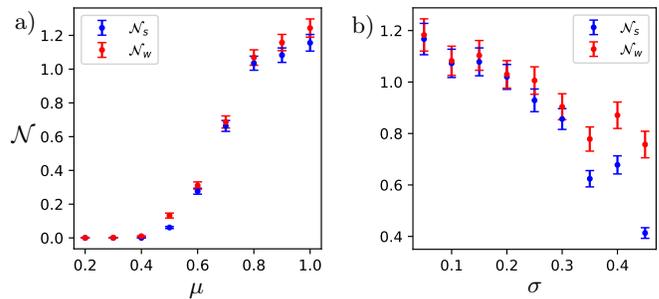
\begin{figure}[t]
\quad
\subfloat{\begin{overpic}[height=1.44in]{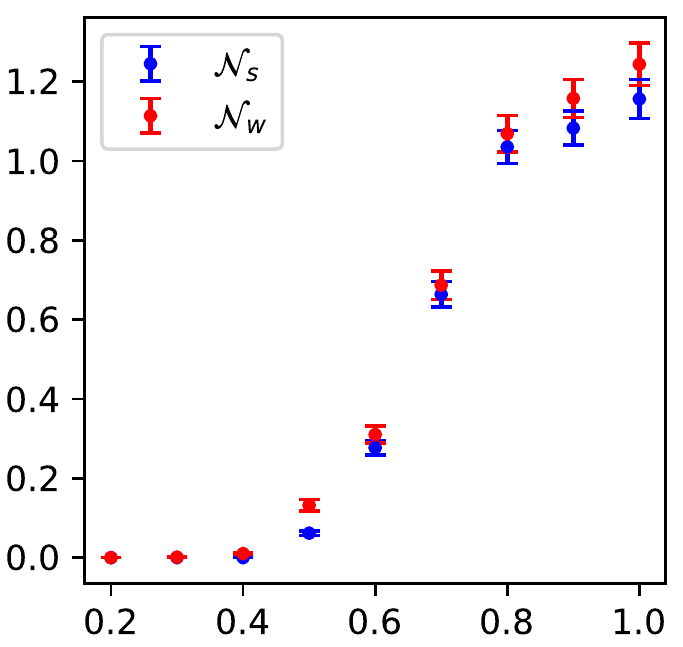}
	\put(-9,46){$\nonmark$}
	\put(52,-6){$\mu$}
	\put(-8, 85){a)}
\end{overpic}}
\qquad
\subfloat{\begin{overpic}[height=1.44in]{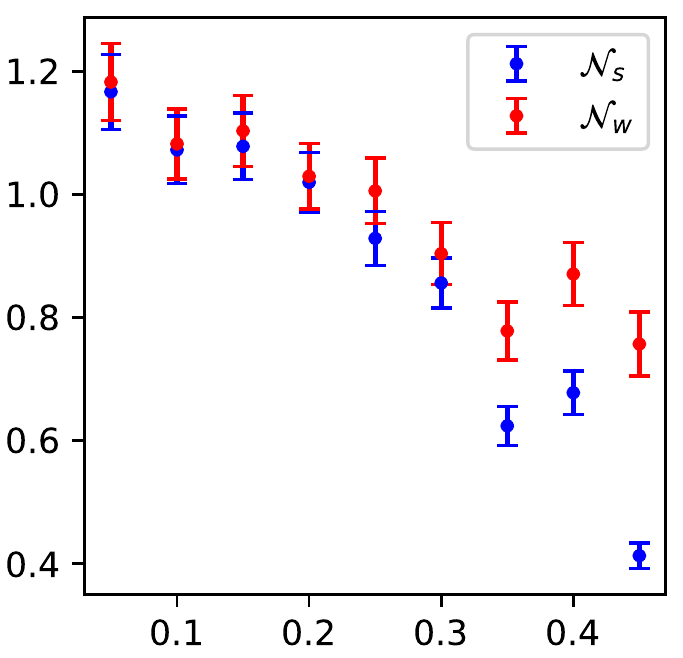}
	\put(52.5,-6){$\sigma$}
	\put(-9, 85){b)}
\end{overpic}}
\caption{(a) Non-Markovianity vs. mean coupling strength $\mu$ for a 10 qubit system coupled to an 11 qubit environment. The plot demonstrates that with 2000 random samples our measures of non-Markovianity can be computed.
(b) Non-Markovianity vs. standard deviation $\sigma$. The plot illustrates that a discrepancy between our measures signifies a varying concentration of non-Markovianity over the space of states.
The error bars denote 90\% confidence intervals.
} \label{fig:numericalDemo}
\end{figure}

We now demonstrate the practicality of our approach and examine $2N + 1$ spin-$1/2$ particles described by the model Hamiltonian,
\begin{equation} \label{eqn:hamiltonian}
    H = \sum_{k=1}^{2N+1} \frac{\omega_k}{2} Z_k + \sum_{k=1}^{2N} J_k \left( X_k Y_{k+1} + Y_k X_{k+1} \right) \,,
\end{equation}
where $X_k, Y_k, Z_k$ denote Paulis on the $k$-th particle, and $\omega_k, J_k$ denote the frequency of the $k$-th particle and the interaction strength between the $k$-th and $(k+1)$-th particle respectively. This type of Hamiltonian models a quantum dot array where Heisenberg interactions decay exponentially over inter-particle distances. Suppose that the particles form a chain of $N$ system qubits coupled to $N+1$ environmental qubits arranged $E-S-E-\dots-E$ where $E$ and $S$ denote the environmental and system qubits respectively. The arrangement corresponds to environmental `defects' that mediate the interactions between system qubits. In Fig.~\ref{fig:numericalDemo} we sample $\omega_k \sim N(0.2, 0.05)$ and $J_k \sim N(\mu, \sigma)$ and consider a total time $T=5$. In Fig. \ref{fig:numericalDemo}.a) we set $\sigma = 0.05$ and vary the mean coupling strength $\mu$. With 2000 samples we determine a relatively small 90\% confidence interval for the non-Markovianity of a system. The size of the system's Hilbert space is $4^8$ times that of the largest space probed numerically by a previous study \cite{Addis2014}. The plot shows that non-Markovian backflow begins at about $\mu = 0.5$. In Fig. \ref{fig:numericalDemo}.b) we vary the standard deviation while fixing $\mu=0.8$. As the variance of the noise grows, the dynamics of each qubit become distinctive. The process deviates from strict non-Markovianity, and the measures diverge.


\begin{figure}[t]
\centering
\begin{overpic}[height=2in]{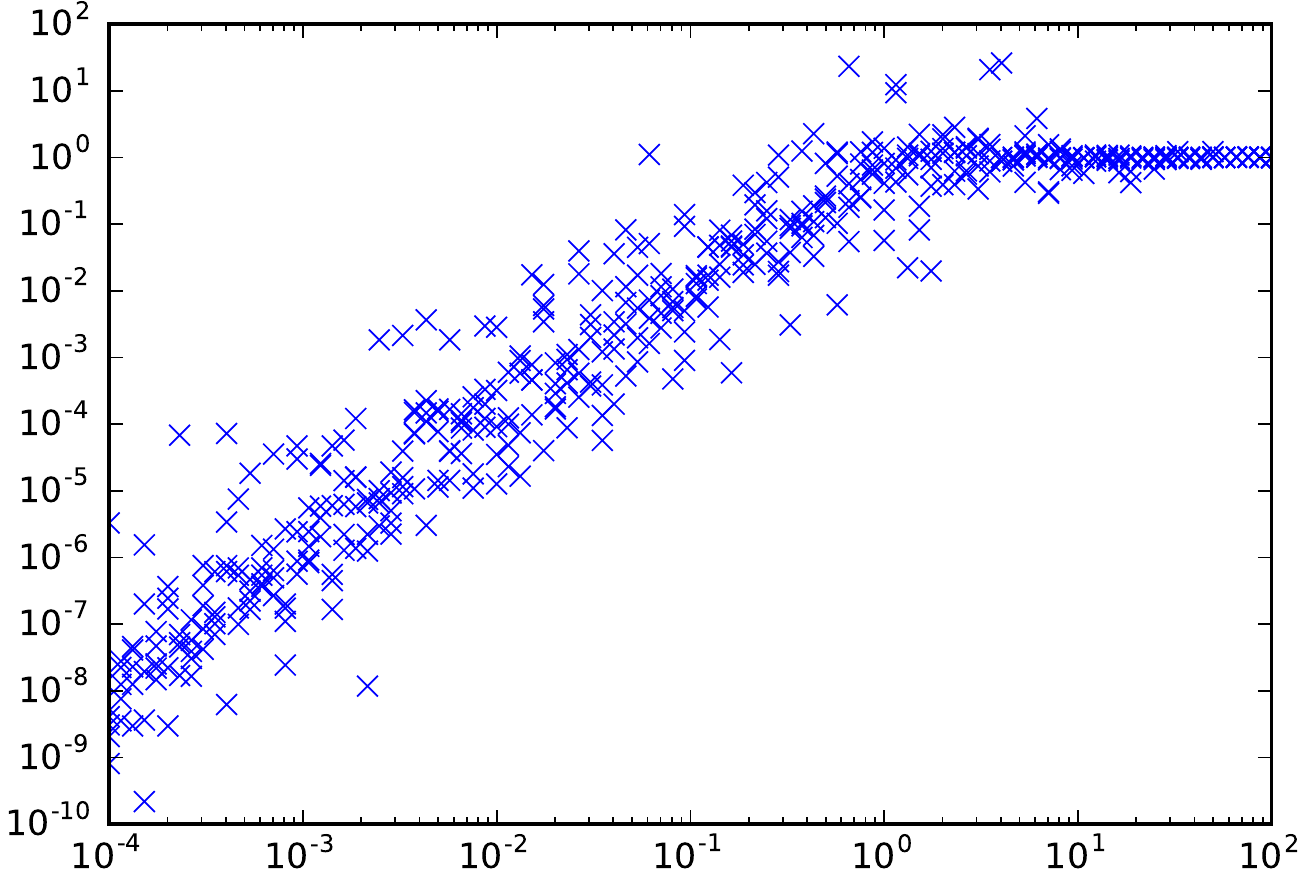}
	\put(-7,35){$\epsilon_\text{rel}$}
	\put(49,-3.2){$h / \tau_c$}
\end{overpic}
\caption{Relative error in approximating $\sigma$ as a function of the finite difference step size $h$. The figure numerically demonstrates that $\tau_c$ describes an upper bound on the order of the step size required for the accurate estimation of $\sigma$.} \label{fig:finiteDiffError}
\end{figure}

The expression for $\sigma$ is experimentally intractable and further requires knowledge of the dynamical map, making it numerically cumbersome. Rather than pursuing an exact solution, we apply the central difference approximation with a step size $h$. We need to bound the timescale of the features in $\sigma$ so that we can determine an upper bound on suitable values for $h$. Without a loss of generality, we consider an interaction Hamiltonian of the form $J = \sum_i J_i \otimes B_i$ where the internal evolution of the environment has been removed by requiring that $\tr(J_i) = 0$ for all $i$. In quantum error-correction theory, the quantity $\lambda = \norm{J}_2$ is a measure of the overall noise-strength \cite{Knill2000}. Thus $\tau_c = 1/\lambda$ is the order of the shortest correlation time scale present in the interaction and provides an upper bound for $h$, and one should usually pick $h \ll \tau_c$.

It is unclear how to bound the second time derivative of the trace distance, and we cannot rigorously bound the local truncation error. Nevertheless, we provide numerical evidence that $\tau_c$ specifies the order of the minimum time resolution needed to apply finite differences. In Fig.~\ref{fig:finiteDiffError} we model 1-3 qubit systems coupled with a 4 qubit environment and employ automatic differentiation to calculate $\sigma$ and compare it with its approximation. The relative error exhibits a quadratic decay that appears as a linear decayx on the log-log plot below $h \sim \tau_c$ and is consistent with the known asymptotic error. The reduction in error occurs as the step size transitions across $\tau_c$ which supports the conjecture that it is an upper bound on $h$. We expect that the decay of $\epsilon_\text{rel}$ is independent of the system size as $\lambda$ identifies the largest singular value in the interaction Hamiltonian.

In summary, we have developed a new approach to measuring the strength of non-Markovianity in a quantum system that builds upon the earlier formalism introduced by \cite{Breuer2009}. By contemplating a toy model, we showed that our approach measures aspects of non-Markovianity that past proposals frequently fail to identify. They are also experimentally and numerically easier to calculate than previous schemes. However, their calculation requires full process tomography and is therefore impractical for large ensembles, such as those encountered in a useful quantum computer. It would be possible to update our approach by identifying efficiently computable quantities that obey a data-processing inequality, i.e., \eqref{eqn:CPmapineq}.

\bibliography{references}
\bibliographystyle{apsrev4-1}

\end{document}